\documentclass{elsart}
\usepackage{epsfig}
\usepackage{rotating}
\begin{document}

\begin{frontmatter}

\leftline{Corresponding Author: {\it Nitesh Soni}}
\leftline{Postal Address:}
\leftline{\it SRF, High Energy Physics Group (Belle Group)}
\leftline{\it Department of Physics,}
\leftline{\it Panjab University - Chandigarh - 160014, India }
\leftline{Phone: \it +91-172-2541741}
\leftline{Fax: \it +91-172-2783336}
\leftline{Email: \it nitesh@puhep.res.in}
\title{ \quad\\[1cm] \Large
  Measurement of Branching Fractions for $B\to {\chi}_{c1(2)} K (K^*)$ at Belle}
\maketitle
\collab{Belle Collaboration}
 \author[Panjab]{N.~Soni}, 
  \author[KEK]{K.~Abe}, 
  \author[TohokuGakuin]{K.~Abe}, 
  \author[KEK]{I.~Adachi}, 
  \author[Tokyo]{H.~Aihara}, 
 \author[BINP]{K.~Arinstein}, 
  \author[Tsukuba]{Y.~Asano}, 
  \author[BINP]{V.~Aulchenko}, 
  \author[ITEP]{T.~Aushev}, 
  \author[Sydney]{A.~M.~Bakich}, 
% ...

\address[BINP]{Budker Institute of Nuclear Physics, Novosibirsk, Russia}
% ...
\address[KEK]{High Energy Accelerator Research Organization (KEK), Tsukuba, Japan}
% ...
\address[ITEP]{Institute for Theoretical and Experimental Physics, Moscow, Russia}
% ...
\address[Panjab]{Panjab University, Chandigarh, India}
% ...
\address[Sydney]{University of Sydney, Sydney, NSW, Australia}
% ...
\address[TohokuGakuin]{Tohoku Gakuin University, Tagajo, Japan}
% ...
\address[Tokyo]{Department of Physics, University of Tokyo, Tokyo, Japan}
% ...
\address[Tsukuba]{University of Tsukuba, Tsukuba, Japan}
% ...
\thanks[NovaGorica]{on leave from Nova Gorica Polytechnic, Nova Gorica, Slovenia}
\begin{abstract}
We have measured the branching fractions for the exclusive decay modes 
$B\to {\chi}_{c1(2)} K(K^*)$ using a $140~{\rm fb}^{-1}$  data sample collected by 
the Belle detector at the KEKB asymmetric-energy $e^+e^-$ collider. The measured branching fractions for
$B^+\to {\chi}_{c1}K^+$,
$B^0\to {\chi}_{c1}K^0$,
$B^0\to {\chi}_{c1}K^{*0}$ and 
$B^+\to {\chi}_{c1}K^{*+}$ decay modes are
$(4.5\pm 0.2\pm0.5)\times 10^{-4}$,
$(3.5\pm 0.3\pm0.5)\times 10^{-4}$,
$(3.1\pm 0.3\pm0.7)\times 10^{-4}$ and
$(4.1\pm 0.6\pm0.9)\times 10^{-4}$, 
respectively, where the first error is statistical and the second error is systematic. We do not observe statistically significant signals for the $B\to {\chi}_{c2}K(K^*)$ decay modes and set upper limits at the $90\%$ confidence level. We also study the helicity distribution for $B\to {\chi}_{c1} K^*$ decay mode and show that
 the longitudinal polarization component is dominant.
\end{abstract}
\end{frontmatter}
\section{Introduction}
Decays of $B$ mesons to two-body final states including a charmonium meson have
played a crucial role in the observation of $CP$-violation in the $B$ meson system~\cite{belle01}.
 These decay modes also provide a sensitive laboratory for studying strong
 interaction (QCD) effects in a heavy meson system. One widely used 
approximation to handle QCD effects in heavy meson decays is 
``factorization", where it is assumed that participating quarks form 
hadrons without subsequent transfer of quantum numbers between the hadrons.
 In the factorization limit, two-body $B$ decays of the type
 $B \rightarrow \chi_{c0} X$ and $\chi_{c2} X$ are suppressed by angular 
momentum and vector current conservation~\cite{suzuki,wir85}. Belle and BaBar have reported the observation of exclusive  $B^{\pm} \rightarrow \chi_{c0} K^{\pm}$~\cite{chic0,chic01,chic02} and the measurement of inclusive $B \rightarrow \chi_{c2} X$ 
decays~\cite{chic2}. A large $B\to \chi _{c0} K$ signal yield can be explained by either factorization violation or rescattering of $D_s^{(*)} D^{(*)}$ in the final state~\cite{colan}{\footnote{ Throughout this paper, when a mode is quoted the inclusion of
 charge conjugate mode is implied.
}}. This mechanism also predicts a branching
 fraction for $B \rightarrow \chi_{c2} K$ comparable to that for $B \rightarrow \chi_{c0} K$. The measurement of new branching fractions of both factorization allowed $B \rightarrow \chi_{c1} K^{*}$ and factorization inhibited 
$B \rightarrow \chi_{c2} K^{*}$ decays provides valuable information for understanding of $\chi _c$ production in $B$ meson decay. In this paper, we report a study of exclusive decays
$B \rightarrow \chi_{c1(2)} K(K^{*})$ for both charged and neutral $B$ mesons,
 where both $\chi_{c1}$ and $\chi_{c2}$ are reconstructed from the same final
 state, $\gamma J/\psi$.

These measurements are based on a $140~{\rm fb}^{-1}$ data sample
 which contains 152 million $B\bar{B}$ pairs,
 collected with the Belle detector~\cite{belle} at the KEKB
 asymmetric-energy $e^+e^-$ (3.5~GeV on 8~GeV) collider~\cite{kekb} operating 
at the $\Upsilon(4S)$ resonance. 
\section{Belle detector}
The Belle detector is a large-solid-angle magnetic spectrometer.
Closest to the interaction point is
a three-layer silicon vertex detector (SVD),
followed by a 50-layer central drift chamber (CDC), 
an array of aerogel threshold \v{C}erenkov counters (ACC),
a barrel-like arrangement of time-of-flight (TOF) scintillation counters, 
 and an electromagnetic calorimeter (ECL) comprised of CsI(T$l$)
 crystals. 
These subdetectors are located inside a superconducting solenoid coil
 that provides a 1.5~T magnetic field. An iron flux-return located outside
 of the coil is instrumented to detect $K^0_L$ mesons and to identify muons. The detector is described in detail elsewhere~\cite{belle}.

\section{Event selection}
Events with $B$-meson candidates are first selected by applying
general hadronic event selection criteria.  These include a requirement on
charged tracks (at least 
     three of them should originate from an event vertex consistent
with the interaction 
     region), a requirement on the reconstructed center-of-mass (CM)
energy ($E^{CM} > 0.2 \sqrt{s}$, where $\sqrt{s}$ is the total CM energy), a
requirement on the longitudinal 
     ($z$-direction) component of the reconstructed CM momentum with
respect to the beam 
     direction ($|p^{CM}_z| < 0.5 \sqrt{s}/c$) and a requirement on
the total ECL energy 
     ($0.1\sqrt{s} < E^{CM}_{ECL} < 0.8\sqrt{s}$) with at least two
energy clusters.
To suppress continuum background, we also require that the
ratio of the second and zeroth Fox-Wolfram moments~\cite{Fox78} be less than
0.5.  To remove charged particle tracks that are poorly
measured or do not come from the interaction region, we require $dz < 5$~cm for all tracks other than those identified as decay daughters
of $K^0_S$, where $dz$ is the track's closest approach to the interaction point along the beam direction.
\section{${\chi}_{c1(2)}$ meson candidates}
The ${\chi}_{c1(2)}$ meson states are reconstructed {\it via} the decay mode
${\chi}_{c1(2)}\to \gamma J/\psi$ with $J/\psi \to \ell^+\ell^-$, where $\ell$ is muon or electron.
 For muon tracks, identification is based on track
penetration depth and the hit pattern in the KLM system~\cite{klm}. Electron tracks
are identified by a combination of $dE/dx$ from the CDC, $E/p$
($E$ is the energy deposited in the ECL and $p$ is the momentum measured
by the SVD and the CDC) and shower shape in the ECL~\cite{ecl}. In order to
recover di-electron events in which one or both electrons have
radiated a photon, the four-momenta of all photons within 0.05
radian of the $e^+$ or $e^-$ directions are included in the invariant
mass calculation. The invariant mass windows are 
$-0.06~(-0.15)~{\rm GeV/c^2}~\le M_{\ell^+\ell^-} - M_{J/\psi} \le 0.036~{\rm GeV/c^2}$ to 
select $J/\psi$ candidates in the $\mu ^+ \mu ^-(e ^+ e ^-)$ channels; these intervals are asymmetric in order to include part of the radiative tails. 
 Vertex- and mass-constrained kinematic fits are then performed for selected $J/\psi$ candidates to improve the momentum resolution. The $\chi_c$ states are reconstructed by
combining a $J/\psi$ candidate with momentum below 2~GeV/c in the CM frame with all
photons of energy greater than 0.060~GeV in the laboratory frame; to suppress $\pi ^0$ we veto photons that, when combined with another photon in the event, satisfy $0.110~{\rm GeV/c^2} \le m_{\gamma\gamma}\le 0.150~{\rm GeV/c^2}$. 
The $\chi_{c1(2)}$ candidates are selected by
requiring the mass difference ($M_{\ell ^+ \ell ^-\gamma} - M_{\ell^+\ell^-}$) to lie
between $0.385~{\rm GeV/c^2}$ and $0.431~{\rm GeV/c^2}$ for
$\chi_{c1}$ and between $0.436~{\rm GeV/c^2}$ and $0.482~{\rm GeV/c^2}$ for $\chi_{c2}$.
 The mass difference resolutions are $9.6~{\rm MeV/c^2}$ and $12.2~ {\rm MeV/c^2}$ at $\chi _{c1}$ and $\chi _{c2}$, respectively. 
 Mass-constrained fits are applied to all selected $\chi _{c1}$ and $\chi _{c2}$ candidates. The ($M_{\ell ^+ \ell ^-\gamma} - M_{\ell^+\ell^-}$) mass distribution is shown in Fig. 1 together with fitted curve~\cite{chic2}.
\section{$K$ or $K^*$ candidates}
The $K^{*0}$ and $K^{*+}$ candidates are reconstructed from four decay
modes: $K^{*0}\to K^+\pi^-$ or $K^0_S\pi^0$, and $K^{*+}\to K^+\pi^0$ or
$K^0_S\pi^+$. 
The kaon likelihood ratio for a track is defined as $\mathcal L(K/\pi) =
\mathcal L(K)/(\mathcal L(K)+\mathcal L(\pi))$. Charged kaons are identified by requiring $\mathcal L(K/\pi)$ to be greater than $0.6$.
The kaon and pion likelihoods ($\mathcal L(K)$~and~$\mathcal L(\pi)$) are obtained by
combining measurements from the TOF and the CDC~$(dE/dx)$ with hit
information from the ACC. 
Tracks which are not identified as either a
kaon or a lepton are treated as charged pion candidates. 
In the reconstruction of the decay mode $B^+\to {\chi}_{c1(2)}K^+$,
all well-measured tracks other than the ones from the $J/\psi$ are considered 
to be kaon candidates.  In order to eliminate the systematic error from particle identification and to retain high efficiency, no particle identification cut for kaons is applied.
The $K^0_S$ candidates are reconstructed from pairs
of oppositely charged pions. The $\pi ^+ \pi ^- $ pair is required to be displaced from the interaction point (IP) by a minimum transverse distance of 0.22 cm for high momentum ($> 1.5 ~{\rm GeV/c}$) candidates and 0.08 cm for those with momentum less than $1.5 ~{\rm GeV/c}$. 
The direction of the pion pair momentum must agree with the direction defined by the IP and the vertex displacement within  0.03 rad for high momentum candidates~($> 1.5 ~{\rm GeV/c}$), within 0.1 rad for  candidates having momentum  between 1.5~GeV/c to 0.5~GeV/c and within 0.3 rad for the remaining low momentum candidates.
The invariant mass of the $\pi^+\pi^-$ pair candidate is required to
satisfy $0.482~{\rm GeV/c^2}~\le M_{\pi^+ \pi^-} \le 0.514~{\rm GeV/c^2}$.
The $\pi^0$ candidates are reconstructed from pairs of photons with energies 
greater than $0.060~{\rm GeV}$. The $\pi^0$
candidates are required to have invariant mass $0.120~{\rm GeV/c^2}~< M_{\gamma\gamma} < 0.150~{\rm GeV/c^2}$.
 A mass-constrained fit is performed to obtain the momentum of the $\pi ^0$. To 
reduce the background from slow pions, we require
the $\pi^0$ momentum to be greater than $0.2~{\rm GeV/c}$ in the CM frame.
\section{$B$ meson reconstruction}
We reconstruct $B$ mesons by combining a ${\chi}_{c1}$ or
${\chi}_{c2}$ candidate with a charged or neutral $K(K^*)$
candidate. We form two independent kinematic variables: the beam
constrained mass $M_{\rm{bc}}\equiv\sqrt{(E_{\rm beam}/c^2)^2 -
((\vec{p}_{\chi_{c1(2)}} + \vec{p}_{K(K^*)})/c)^2}$, and $\Delta E \equiv
(E_{\chi_{c1(2)}} + E_{K(K^*)}) - E_{\rm beam}$, where $E_{\rm beam}$ is
the beam energy, and $p_{\chi_{c1(2)}}$, $E_{\chi_{c1(2)}}$, $p_{K(K^*)}$
and $E_{K(K^*)}$ are momenta and energies in the CM frame of the
$\chi_c$ states and the $K(K^*)$ mesons, respectively. 
The $B$-meson signal window is defined as $5.27~{\rm GeV/c^2} \leq
M_{\rm {bc}} \leq 5.29~{\rm GeV/c^2}$ for all channels and 
the ranges for $\Delta E$ are $\pm$ 0.025$~{\rm GeV}$ for $B\to {\chi}_{c1(2)}K(K^0_S)$, and 
$\pm$~0.040 $~{\rm GeV}$ for $B\to {\chi}_{c1(2)}K^*(K^+\pi^-$ or $K^0_S\pi^+$). 
A larger range ($-0.050~{\rm GeV} \le {\Delta E} \le 0.030~{\rm GeV}$) is used for $B\to {\chi}_{c1(2)}K^{*+}(K^+\pi^0$ or 
$K^0_S \pi^0)$ to accommodate the wider $\Delta E$ distribution that results from energy leakage in the calorimeter for $\pi ^0$ modes.
\section{Signal yield and background estimation}
To study various sources of background, we use 5.5 million $B\rightarrow \psi$
inclusive Monte Carlo events generated using the event generator
EvtGen~\cite{evtgen} and a GEANT3-based program~\cite{geant} to simulate
the Belle detector response. This Monte Carlo sample contains decays of
 the type $B\to k$, where $k$ is a three-body final state, e.g.
 $\chi _{c1} K\pi$ or $J/\psi K\pi$ etc., or two-body decays, e.g. $J/\psi K $ or $J/\psi K^*$ or $\chi _{c1(2)} K(K^*)$ etc. Based on this Monte Carlo study, 
the backgrounds for $B\to{\chi}_{c1(2)}K(K^0_S)$ decay modes can be categorized as $J/\psi$ inclusive
and feed-across
backgrounds. The $J/\psi$ inclusive background occurs due to a
true $J/\psi$ candidate accidentally combining with an unrelated $\gamma$ to fake a 
${\chi}_{c1(2)}$. 
Feed-across background is due to one
mode being reconstructed as a different mode because of the similarity in
event kinematics. For $B\to {\chi}_{c1(2)}K^*$ decay modes, there is an additional
background due to non-resonant decays of $B$ mesons, i.e. $B\to \chi _{c1(2)} K \pi$. Non-resonant background peaks at the same place as
signal in the $M_{\rm {bc}}$ and $\Delta E$ distributions, but does not
peak in the $K\pi$-mass distribution.  We thus extract the number of
signal events from a fit to the $M_{\rm {bc}}$ distribution for $B\to
{\chi}_{c1(2)}K(K^0_S)$ decay modes but use the $K\pi$-mass distribution for $B\to
{\chi}_{c1(2)}K^*$ decay modes. Each component of background has been
estimated separately as discussed below.

The $J/\psi$ inclusive background is modeled by
the sidebands in the ($M_{\ell ^+ \ell ^-\gamma} - M_{\ell^+\ell^-}$) mass distribution of data
 and  scaled according to the fitted area of the background under
the ${\chi}_{c1(2)}$ signal region. The sidebands are defined as 0.2 ${\rm GeV/c^2}$ to 0.3 ${\rm GeV/c^2}$ and 0.55 ${\rm GeV/c^2}$ to 0.60 ${\rm GeV/c^2}$ 
in the ($M_{\ell ^+ \ell ^-\gamma} - M_{\ell^+\ell^-}$) mass distribution. This model  includes a
contribution from the combinatorial $\ell^+\ell^-$ background under the $J/\psi$
peak (``fake $J/\psi$ candidates''), which is not present in the
$\psi$ inclusive Monte Carlo sample; this component forms about 5$\%$ of the $J/\psi$
inclusive background.
As a systematic check, the $\psi$ inclusive Monte Carlo has also been used to model the $J/\psi$ inclusive
background and the contribution of the $J/\psi $ inclusive background decays have 
been obtained by excluding the feed-across and non-resonant background decays 
in the $\psi$ inclusive Monte Carlo sample. A small difference in the fit between 
the two models has been included in the systematic error. 

For each decay mode $i=1,......12$, the feed-across background due to 
other $\chi_{c1(2)} K(K^*)$ decay
modes $j=1,.......12$ with $j \neq i$ is estimated by taking the
 appropriate Monte Carlo
histograms and normalizing them according to the signal
yields observed in the modes $j$. The total feed-across
background for mode $i$ is then fixed in the fit. The
fitting procedure for all twelve $\chi_{c1(2)}K(K^*)$
modes is repeated, using the signal yields from the
previous iteration as input, until we find a stable result. 

The non-resonant background shape is obtained from the $\psi$~inclusive Monte Carlo sample. Due to low statistics and
 similar shapes of the $J/\psi$ inclusive and non-resonant
background, it is not possible to float simultaneously the normalizations of both backgrounds in the fit to data. 
We have little knowledge about the amount of non-resonant background, so when 
fitting the $K\pi$-mass distribution the 
normalization of the $J/\psi$ inclusive background is fixed, while it is floated 
for non-resonant background. The contributions from non-resonant background decays 
have been obtained from the $\psi$ inclusive Monte Carlo sample by excluding the 
feed-across and $J/\psi$ inclusive background decays.

The $M_{\rm {bc}}$ distributions for $B\to {\chi}_{c1(2)}K(K^0_S)$ decay modes
are fitted with a Gaussian to describe the signal region. The width and mean 
value of the Gaussian are fixed in the fit using Monte Carlo values that are calibrated
 by the $B^+ \rightarrow \chi_{c1} K^+$ decay mode in data because this mode has the highest
 statistics among the decay modes considered. 
The background shapes for the $J/\psi$ inclusive and feed-across backgrounds have been
modeled as discussed above. The normalization parameter for the feed-across background histogram is fixed while it is floated for the $J/\psi$ inclusive background histogram.
Figure 2 displays the fit to the $M_{\rm {bc}}$ distributions for the
(a) $B^+\to {\chi}_{c1}K^+$,
(b) $B^+\to {\chi}_{c2}K^+$,
(c) $B^0\to {\chi}_{c1}K^0_S$, and
(d) $B^0\to {\chi}_{c2}K^0_S$ decay modes. The signal yields extracted from these distributions are listed in Table 1. For the $B\to {\chi}_{c2}K(K^0_S)$ mode,
no statistically significant signal is seen. As a check, the signal yields have also been obtained by fitting the $\Delta E$ distributions for the $B\to {\chi}_{c1(2)}K(K^0_S)$ decay modes and are found to be consistent with the signal yields obtained from the fits to the $M_{\rm bc}$ distributions.

A relativistic P-wave Breit-Wigner function~\cite{breit} is used to fit the $K^*(892)$ signal
region in the $K\pi$-mass distributions for the $B \to {\chi}_{c1(2)}K^*$ decay modes.
For $K^*(892)$, the width and peak position of the P-wave Breit-Wigner function are found to be consistent between data and Monte Carlo in the  $B\to J/\psi K^*(892)$ decay 
modes~\cite{itoh}; therefore, those parameters are fixed to the PDG values in the fit~\cite{Hag}. The shapes of the
$J/\psi$ inclusive, feed-across and non-resonant backgrounds are modeled as
explained previously. 
There may be some structure in the $K_2^*(1430)$~ region in the $K\pi$-mass distributions for the $B\to {\chi}_{c1(2)}K^*$ decay modes. To accommodate this possibility, we have included a relativistic D-wave Breit-Wigner function in the fits to the $K\pi$-mass distributions. The mean and width of the D-wave Breit-Wigner function are fixed to the nominal PDG values.
Figure 3 shows the fits to the $K\pi$-mass distribution for the
(a) $B^0\to {\chi}_{c1}K^{*0}(K^+\pi^-)$,
(b) $B^0\to {\chi}_{c1}K^{*0}(K^0_S\pi^0)$,
(c) $B^+\to {\chi}_{c1}K^{*+}(K^+\pi^0)$, and
(d) $B^+\to {\chi}_{c1}K^{*+}(K^0_S\pi^+)$ decay modes.
The fits to the $K\pi$-mass distributions for
(a) $B^0\to {\chi}_{c2}K^{*0}(K^+\pi^-)$,
(b) $B^0\to {\chi}_{c2}K^{*0}(K^0_S\pi^0)$,
(c) $B^+\to {\chi}_{c2}K^{*+}(K^+\pi^0)$, and
(d) $B^+\to {\chi}_{c2}K^{*+}(K^0_S\pi^+)$ decay modes are presented in Fig. 4.
The signal yields extracted from these distributions are listed in Table 1. For the $B^0\to {\chi}_{c1}K^{*0}(K^0_S \pi^0)$ and $B\to {\chi}_{c2}K^{*}$~decay modes, no statistically significant signals are seen. 
The signal yields have also been checked by fitting the $M_{\rm bc}$ and $\Delta E$ distributions for the $B\to {\chi}_{c1(2)}K^*$ decay modes. For the $M_{\rm bc}$ and $\Delta E$ distributions a $K^*$ is identified if the absolute difference between the invariant mass of an identified $K\pi$ pair and nominal $K^*$ mass is less than $75~{\rm MeV/c^2}$. The signal yields obtained from the fits to the $M_{\rm bc}$ and $\Delta E$ distributions are found to be consistent with the signal yields obtained from the fits to the $K\pi$-mass distributions.
\section{Helicity study for $B\to \chi _{c1} K^*$ decays}
A $P\to VV$ decay may include three possible polarization 
amplitudes: two transverse and one longitudinal. The contributions  of these 
amplitudes can be obtained from the helicity distribution of the 
$K^*$ ($\cos \theta _{K^*}$) where $\theta _{K^*}$ is defined as the angle between the $K^*$ momentum and the direction opposite to the $B$ momentum in the $K^*$ rest frame.
 The transverse polarization
 amplitudes have a $\sin ^2 \theta _{K^*}$ dependence while the longitudinal amplitude has 
a $\cos ^2 \theta _{K^*}$ dependence. We have analyzed the helicity distributions
 for different $K\pi$-mass regions. The efficiency-corrected $\cos \theta _{K^*}$ distribution for the signal region 
($0.821~{\rm GeV/c^2} < {\rm M_{K\pi}} < 0.971~{\rm GeV/c^2}$)
is fitted with the function 
$f(\cos \theta _{K^*}) = a+3b\cos ^2 \theta _{K^*}+2c{\sqrt {3ba}}\cos \theta _{K^*}$. The fitted value of $c$ ($0.01\pm 0.10$) is consistent with zero, which shows that the interference with the scalar amplitude is negligible. The helicity distribution 
for the background obtained from sideband region (${\rm M_{K\pi}} < 0.8~{\rm GeV/c^2}$)
 is flat except for a peak in the last bin due to the $J/\psi$ inclusive background  (Fig. 5(a)); this distribution is significantly different from the helicity distribution for the signal region as shown in Fig. 5(b). We perform a fit to the efficiency-corrected $\cos\theta_{K^*}$
distributions for the signal and sideband regions with the function
$F(\cos\theta_{K^*})=A+3B\cos^2\theta_{K^*}$
and obtain 
$A_{\rm sig}$($A_{\rm sb}$) and $B_{\rm sig}$($B_{\rm sb}$)
for the $K^*$ signal (sideband) region.
With these parameters, the background subtracted net $K^*$ helicity angle distribution is 
extracted with corresponding parameters $A^*$ and $B^*$ by estimating 
$A^* = A_{\rm sig} - S A_{\rm sb}$ and 
$B^* = B_{\rm sig} - S B_{\rm sb}$, 
where $S$ is the scale factor to subtract non-resonant
$K\pi$ contributions and other backgrounds. 
Thus, based on $A^*$ and $B^*$, we obtain the longitudinal polarization of $K^*$.
Using the longitudinal polarization amplitude ${\mathcal A}_{\rm L}$ and the total 
amplitude ${\mathcal A}_{\rm tot}$, the longitudinal polarization parameter is 
expressed as $\eta = { |{\mathcal A_{\rm L}}|^2 \over |{\mathcal A_{\rm tot}}|^2}$ 
which is also expressed in terms of $A^*$ and $B^*$; 
$\eta = { B^* + {A^* \over 3} \over {A^* + B^*}}$.
 It is also observed that the difference in fit with or without inclusion of the last peaking bin of the $\cos \theta _{K^*}$ distribution for the background region is small and the error assigned to $\eta$  takes into account this effect. 
To estimate the systematics due to the background shape uncertainty, we obtain the value of $\eta$
 by using the upper sideband region ($1.1~{\rm GeV/c^2} < \rm M_{K\pi} < 1.3~{\rm GeV/c^2}$)
for a background fit.
The difference in values obtained using the two different sideband regions
 is found to be 4.6$\%$, which is treated as a systematic error. To estimate the influence of a possible interference 
term ($\cos\theta_{K^*}$) on the
polarization parameter value, we perform fits taking the signal events from left
and right sides of the $K^*$ peak i.e. $(0.821~{\rm GeV/c^2} < \rm M_{K\pi} < 0.896~{\rm GeV/c^2})$ and $(0.896~{\rm GeV/c^2} < \rm M_{K\pi} < 0.971~{\rm GeV/c^2})$.
In spite of the observed nonzero values of the scalar interference term in these distributions, 
the change in longitudinal polarization is found to be within 6.8$\%$, which is taken as a systematic error. These two errors have been combined in quadrature to give a total systematic error of 8.2$\%$ on polarization parameter ($\eta$).
 Hence the final value of longitudinal polarization parameter $\eta$ from this analysis is obtained as $\eta = 0.87 \pm 0.09 \pm 0.07$.
From this value of $\eta$ we  
conclude that the $B\to \chi _{c1}K^*$ decay is dominated by longitudinal 
polarization with a lower limit for the longitudinal polarization of $ 0.69$ at the 90$\%$ confidence level.
 This mode has a larger longitudinal
 polarization than $B\to J/\psi K^*$~\cite{itoh,Jpsibabar}.
\section{Measurement of branching fractions}
The branching fraction for each $B\to {\chi}_{c1(2)}K(K^*)$ decay mode
 is estimated by dividing the observed signal yield by the
 reconstruction efficiency, the number of $B{\bar B}$ events in the
 data sample and the daughter branching fractions. We have used the
 daughter branching fractions published in PDG 2004~\cite{Hag}. Equal
 production of neutral and charged $B$ meson pairs in
 $\Upsilon ({\rm 4S})$ decay is assumed. Reconstruction efficiencies are
 estimated by applying the same selection criteria to 20,000 signal
 Monte Carlo events. The efficiencies for each final state are shown in Table 1. Here, the branching fraction of $K^0 \rightarrow K^0_S \rightarrow \pi^+\pi^-$ is not included in the efficiency calculation.
In the fits to data, the width of the ${\chi}_{c2}$ is constrained 
to be 10$\%$ larger than the width of 
the ${\chi}_{c1}$ as expected from Monte Carlo simulation and the photon
resolution. The effect of this assumption is included in the systematic errors. To calculate upper limits (U.L.), we add the statistical and
                systematic errors in quadrature, and assume that the
                combined error behaves as a Gaussian. We then follow the
                prescription of Feldman and Cousins~\cite{Feld}.
The branching fractions are summarized in
Table 1.
The first errors are statistical and the second errors are systematic. 
The statistical significance ($\Sigma$) of the
signal in terms of the number of standard deviations is calculated as
$\sqrt{-2ln(\mathcal L_0/\mathcal L_{max})}$, where $\mathcal L_{max}$ and $\mathcal L_0$ denote the
maximum likelihood with the nominal signal yield and with the signal
yield fixed at zero, respectively. If a decay mode is reconstructed 
in two final states, e.g. $B^+\to{\chi}_{c1}K^{*+}(K^0_S\pi^+)$ and $B^+\to {\chi}_{c1}K^{*+}(K^+\pi^0)$,
the two branching fraction measurements are averaged, weighted by the statistical and systematic errors. Comparison between our measurement and those at BaBar~\cite{b1,b2} is also shown in Table 3.
\section{Systematic uncertainties}
Various sources of systematic uncertainty are summarized in
Table~2. The uncertainty on the tracking efficiency is
estimated to be 1.5$\%$ for $e$ and $\pi$ tracks, and 1.1$\%$ for $\mu $ and $K$ tracks.
The pion tracks from $K^0_S$ decay are
from a displaced vertex and thus have a larger systematic
error; we include 2$\%$ per track uncertainty for pion tracks from $K^0_S$~ decays.
The errors on electron and muon identification are 1.6$\%$ and 2.2$\%$, respectively. The kaon
identification error is estimated to be $1.9\%$.\\

Uncertainties on $\gamma$ and $\pi^0$ reconstruction efficiency are 2$\%$
and 4$\%$, respectively.
The uncertainty in the $K^0_S$ selection efficiency is checked by
comparing yields for a sample of high momentum $K^0_S \to \pi^+ \pi^-$
decays before and after applying the $K^0_S$ selection criteria. The
efficiency difference between data and Monte Carlo simulation is less
than 1.0$\%$.\\ 

All modes have a systematic uncertainty due to ${\chi}_{c1(2)}\to \gamma J/\psi$
and $J/\psi\to \ell^+ \ell^-$ branching fractions; the uncertainties for these
modes are taken from the PDG~\cite{Hag} and added in quadrature.\\

The detection efficiency is found to have so small dependence as
a function of decay angles that the integrated interference term
among amplitudes with different quantum number of the $K\pi$-system gets
almost cancel. Therefore, this effect on the branching fraction
is estimated to be less than 1.3$\%$. 
However, the integrated amplitudes should be added coherently for the non-resonant $K\pi$ background of the same quantum numbers.
To estimate possible systematic errors due to interference between the signal and background, we fit the $K\pi$-mass 
distribution to a function which assumes a ``coherent'' background contribution. We define this function as 
\begin{center}
\begin{equation}
F(m_{K\pi}) = | \sqrt{N_s}S(m_{K\pi}) + {\sqrt{(N_b a_c)}}e^{i \phi}B_g(m_{K\pi})|^2 + N_b(1-a_c)B_g^2(m_{K\pi})
\end{equation}
\end{center}
where
\begin{center}
\begin{equation}
S(m_{K\pi}) = \frac {m_{K^*}\Gamma(m_{K\pi})}{(m^2_{K\pi} - m^2 _{K^*}) + i m_{K^*}\Gamma (m_{K\pi})}
\end{equation}
\end{center}
is a relativistic P-wave Breit-Wigner function with $m_{K^*} = 892~{\rm MeV/c^2}$, $B_g(m_{K\pi})$ is the phase space background function, $N_s$ and $N_b$ are the numbers of events for signal and background, respectively, and $a_c$ is the fraction of coherent background with a varied phase $\phi$ relative to the signal. 
The maximum value of the coherent background fraction is taken from the helicity 
analysis to be $a_c = 0.6$. By varying the phase and relative fraction of coherent background 
within error in this fit, we find that the maximum difference in the signal yield due to the possible interference term is 15$\%$. This difference is included as a systematic error and listed as ``Inter.'' in Table 2. We assume that this difference is same for all $K^*$ channels; it is determined from the highest statistics $B\to \chi _{c1} K^*$ mode. 

In this study we have assumed that the background shape is described by the phase space function. We also fit the same distribution by using different background shapes. In one set of fits, a threshold function~\cite{itoh} is used to describe the background shape and in another set the 
function $B_g(m_{K\pi})(a_1+a_2 P_K+a_3 P_{\chi _{c1}})$  is used to describe the background shape; here $B_g(m_{K\pi})$ is the phase space background function, $a_1$, $a_2$, $a_3$ are the free parameters in the fit, $P_K$ is the momentum of $K$ in the $K^*$ rest frame and $P_{\chi _{c1}}$ is the $\chi _{c1}$ momentum in the $B$ rest frame. The maximum difference in the signal yield obtained by using different background shapes is found to be 10.8$\%$ and is taken as an additional systematic error listed as background shape (``Bkg. Shape'') in Table 2. This difference is also determined from the highest statistics $B\to \chi _{c1} K^*$ mode and assumed to be the same for all $K^*$ modes. 

The detection efficiency can depend on the various values of polarization amplitudes for $B\to \chi _{c1(2)} K^*$ decay modes.
 To check this effect, the exclusive Monte Carlo sample of $B\to \chi _{c1} K^*(K^+ \pi ^-)$ decay with various values of the polarization amplitudes is generated. The maximum difference in detection efficiency is estimated to be 5.1$\%$ and treated as systematic error. This difference is assumed to be the same for all $K^*$ modes.

The yield measurements are taken from fits where the mean and width of the signal shape 
are fixed. The corresponding uncertainty is estimated by varying the means and widths
in turn by $\pm 1\sigma$, and repeating the fit.
 The percentage change in the signal yield is presented in Table 2. 
The difference between two background fitting models (i.e. using the $J/\psi$ inclusive background shapes from data or from the $\psi$ inclusive Monte Carlo) has also been included as a systematic error  listed as background error (``Bkg. Error'') in Table 2.

For the ${\chi}_{c1}$ mode, the error on the efficiency correction due to 
the ${\chi}_{c1}$ width is based on the uncertainty of the fitted width in 
data. For the ${\chi}_{c2}$ mode, there are additional effects due to the assumptions in the fit:  $\sigma(\chi _{c2}) : \sigma(\chi _{c1})~=~1.1 : 1$ (estimated in an alternative fit where the ${\chi}_{c1}$ and ${\chi}_{c2}$ widths float independently) and the uncertainty on the correction factor for the feed-across from ${\chi}_{c1}$. The resulting error estimates are combined in quadrature, and listed as an efficiency correction (``Eff. Corr.'') in Table~2.\\

The uncertainty on the number of
 $B{\bar B}$ events, which is common to all modes, is 1$\%$; this is
negligible compared to the other systematic errors.
\section{Conclusion}                               
We have measured the branching fractions for the $B\to
{\chi}_{c1(2)}K(K^*)$ decay modes using 152 million $B{\bar B}$ events. 
Although inclusive $\chi _{c2}$ signals in $B$ decay have been observed before~\cite{chic2}, we do not find any significant signals for the exclusive two-body $B\to \chi_{c2} K(K^*)$ decay modes and set upper limits on the branching fractions at the 90$\%$ confidence level which are 
an order of magnitude smaller than those for $B\to {\chi}_{c1}K(K^*)$. 
The helicity distribution for the $B\to \chi _{c1} K^*$ decay mode 
shows that the longitudinal polarization component is dominant.
\section{Acknowledgments}
We thank the KEKB group for the excellent operation of the
accelerator, the KEK cryogenics group for the efficient
operation of the solenoid, and the KEK computer group and
the National Institute of Informatics for valuable computing
and Super-SINET network support. We acknowledge support from
the Ministry of Education, Culture, Sports, Science, and
Technology of Japan and the Japan Society for the Promotion
of Science; the Australian Research Council and the
Australian Department of Education, Science and Training;
the National Science Foundation of China under contract
No.~10175071; the Department of Science and Technology of
India; the BK21 program of the Ministry of Education of
Korea and the CHEP SRC program of the Korea Science and
Engineering Foundation; the Polish State Committee for
Scientific Research under contract No.~2P03B 01324; the
Ministry of Science and Technology of the Russian
Federation; the Ministry of Higher Education, Science and Technology of the Republic of Slovenia;  the Swiss National Science Foundation; the National Science Council and
the Ministry of Education of Taiwan; and the U.S.\
Department of Energy.

\begin{figure}[!tbh]
\begin{center}
   \begin{tabular}[p]{c}       
    \includegraphics[width=0.7\textwidth]{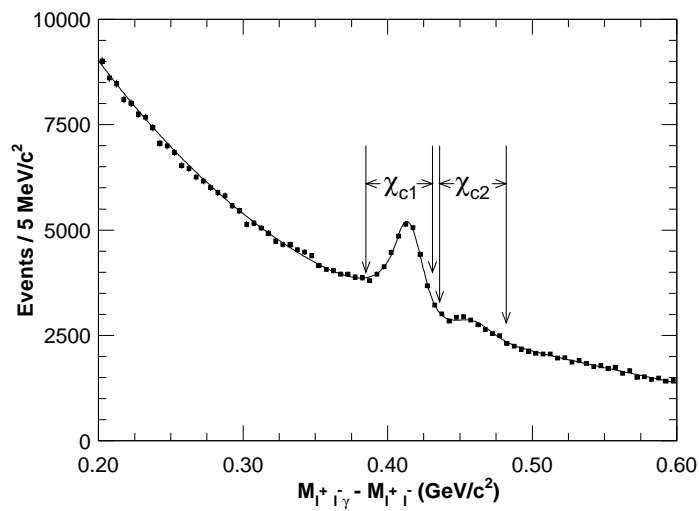}   
\end{tabular}

\caption{The ($M_{\ell ^+ \ell ^-\gamma} - M_{\ell^+\ell^-}$) mass difference distribution with the selected $\chi _{c1}$ and $\chi _{c2}$ mass regions.}
\end{center}
\end{figure}
\begin{figure}[!tbh]
\begin{center}
   \begin{tabular}[p]{c}       
    \includegraphics[width=1.0\textwidth]{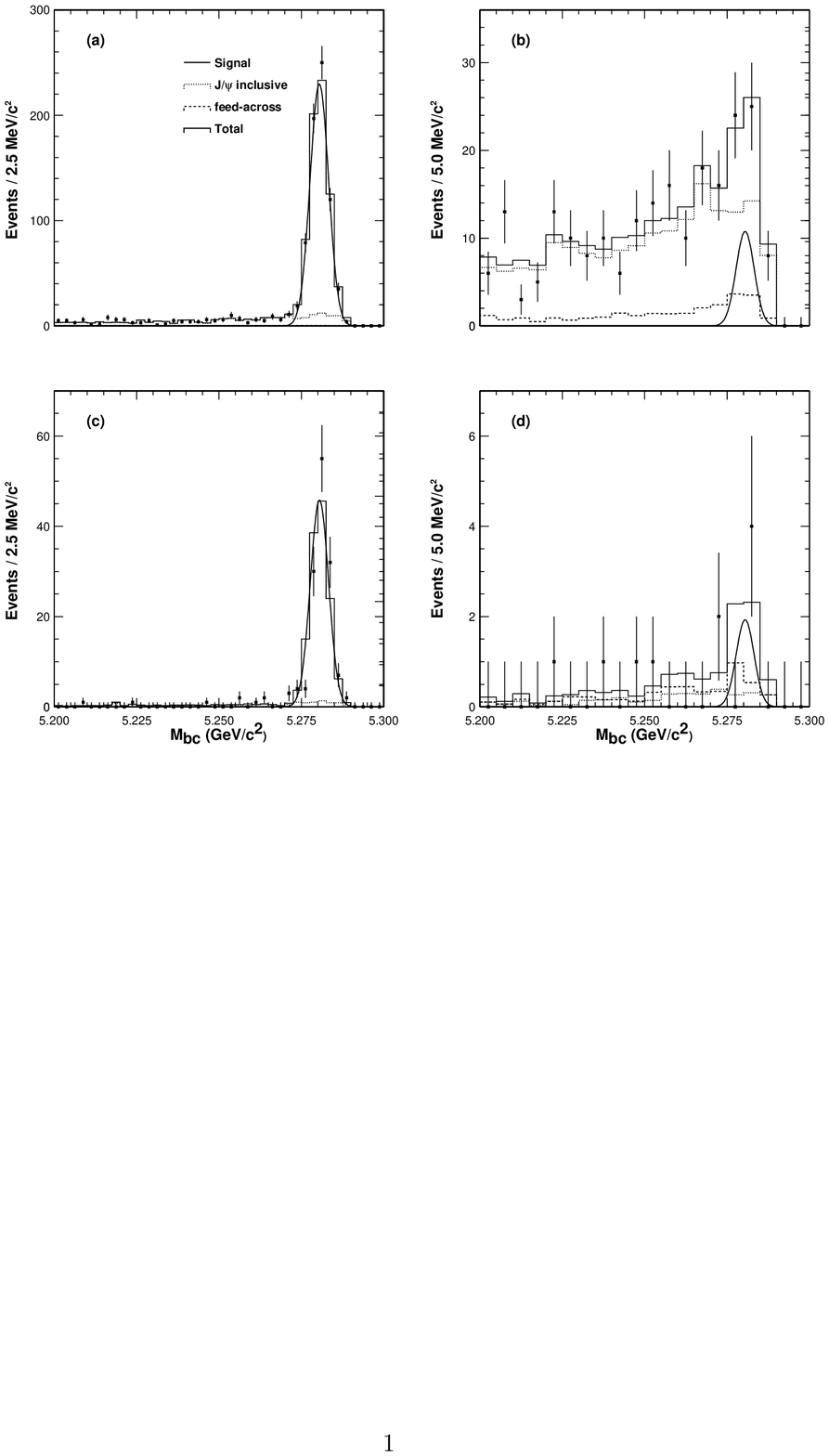}   
\end{tabular}

\caption{The $M_{\rm {bc}}$  distribution for (a) $B^+\to {\chi}_{c1}K^+$, (b) $B^+\to {\chi}_{c2}K^+ $,
(c) $B^0\to {\chi}_{c1}K^0_S$ and (d) $B^0\to {\chi}_{c2}K^0_S$ decay modes, respectively, together with results of the fit for the signal region and contributions due to $J/\psi$ inclusive and feed-across backgrounds.}
\end{center}
\end{figure}
\begin{figure}[!tbh]
\begin{center}
   \begin{tabular}[p]{c}       
    \includegraphics[width=1.0\textwidth]{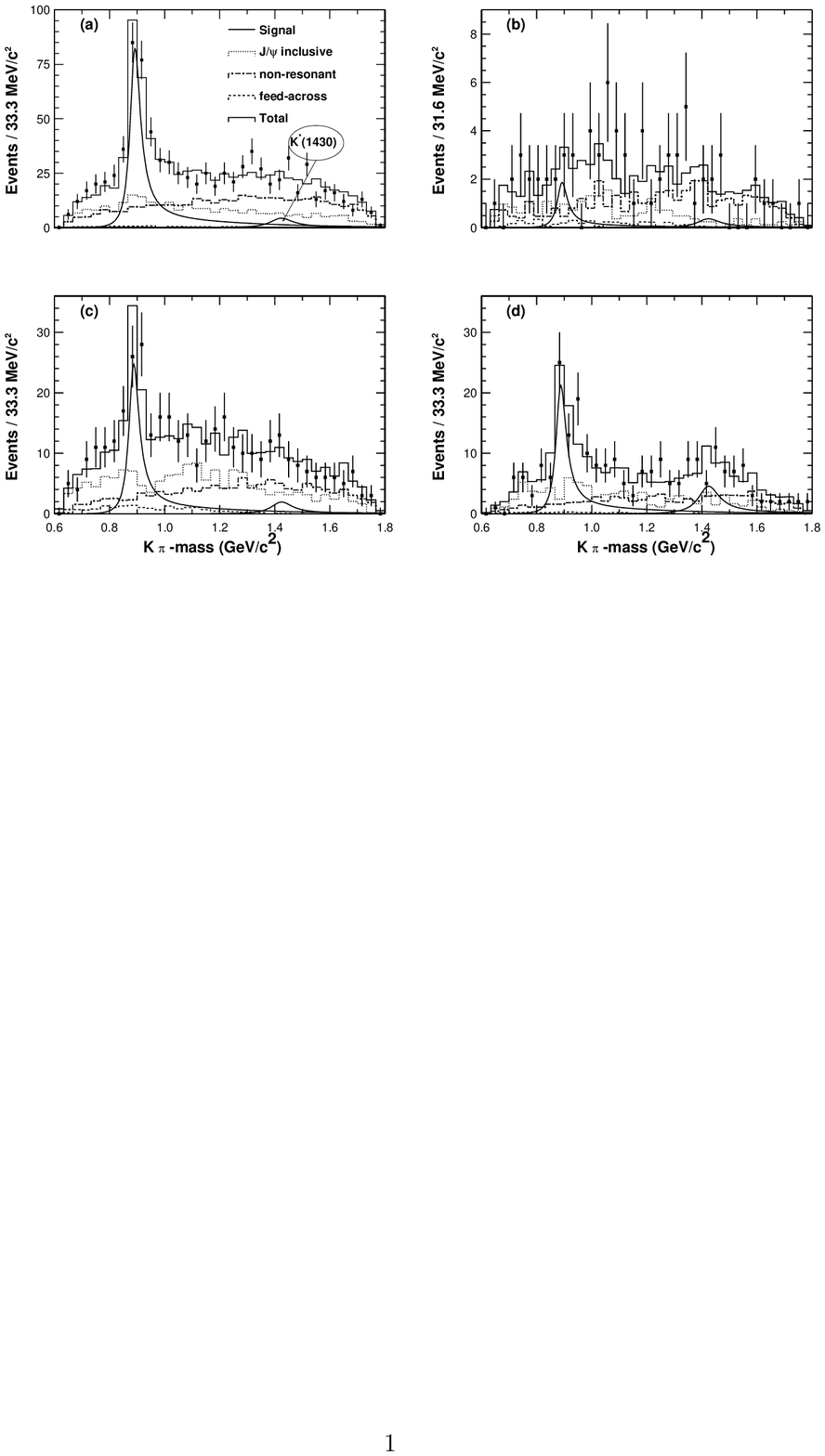}  
\end{tabular}
\caption{The $K\pi$-mass distribution for (a) $B^0\to {\chi}_{c1}K^{*0}(K^+\pi^-)$, 
(b) $B^0\to {\chi}_{c1}K^{*0}(K^0_S\pi^0)$, 
(c)  $B^+\to {\chi}_{c1}K^{*+}(K^+\pi^0)$, and  (d) $B^+\to {\chi}_{c1}K^{*+}(K^0_S\pi^+)$ decay modes, respectively, together with results from the fit to $K^*$ signals and contributions of backgrounds due to $J/\psi$ inclusive, non-resonant and feed-across backgrounds.}
\end{center}
\end{figure}
\begin{figure}[!tbh]
\begin{center}
   \begin{tabular}[p]{c}
    \includegraphics[width=1.0\textwidth]{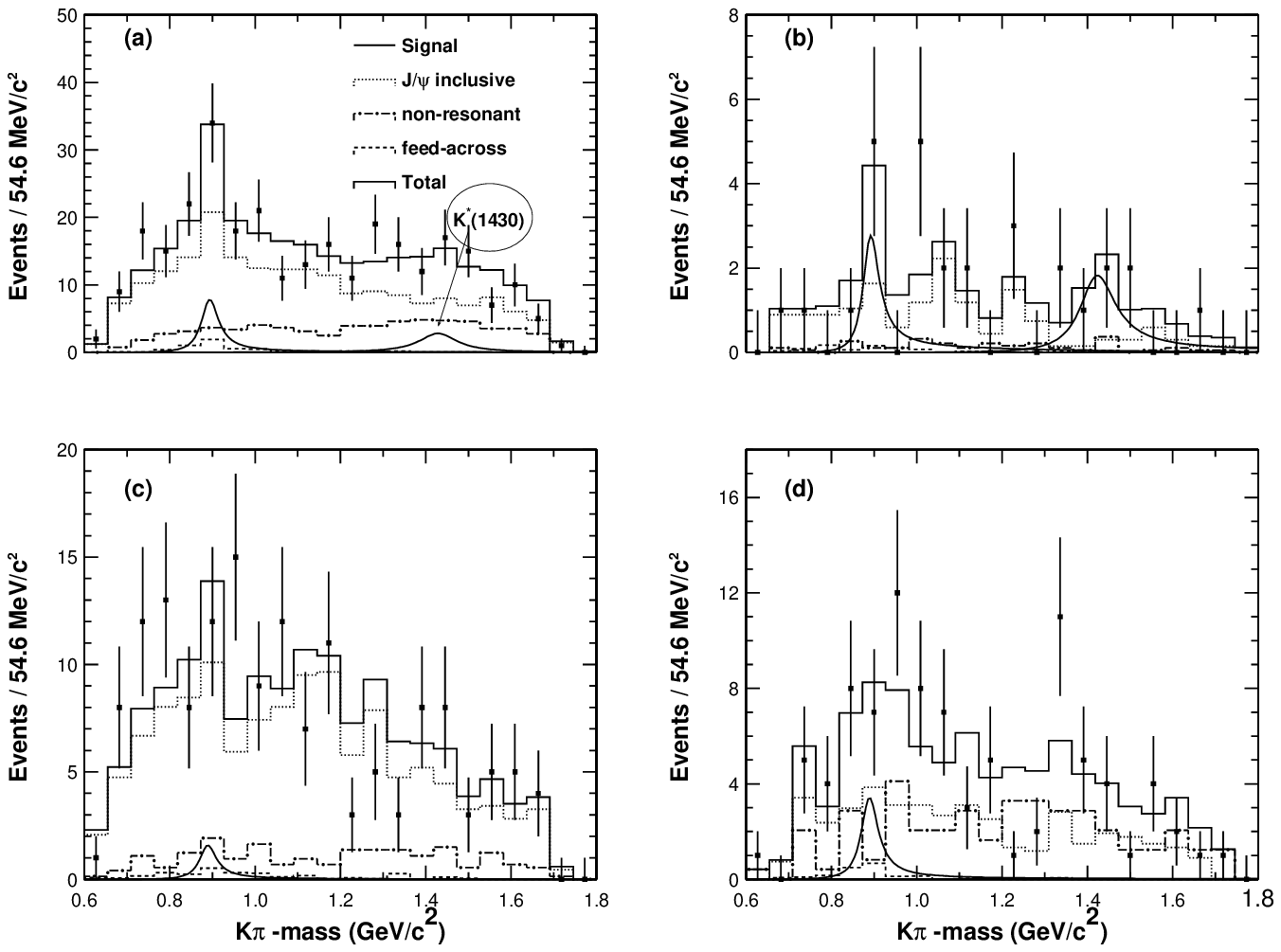}  
\end{tabular}
\caption{The $K\pi$-mass distribution for (a) $B^0\to {\chi}_{c2}K^{*0}(K^+\pi^-)$, 
(b) $B^0\to {\chi}_{c2}K^{*0}(K^0_S\pi^0)$, 
(c)  $B^+\to {\chi}_{c2}K^{*+}(K^+\pi^0)$, and  (d) $B^+\to {\chi}_{c2}K^{*+}(K^0_S\pi^+)$ decay modes, respectively, together with results from fit to $K^*$ signals and contributions of backgrounds due to $J/\psi$ inclusive, non-resonant and feed-across backgrounds.}
\end{center}
\end{figure}
\begin{figure}[!tbh]
\begin{center}
   \begin{tabular}[p]{c}
       \includegraphics[width=1.0\textwidth]{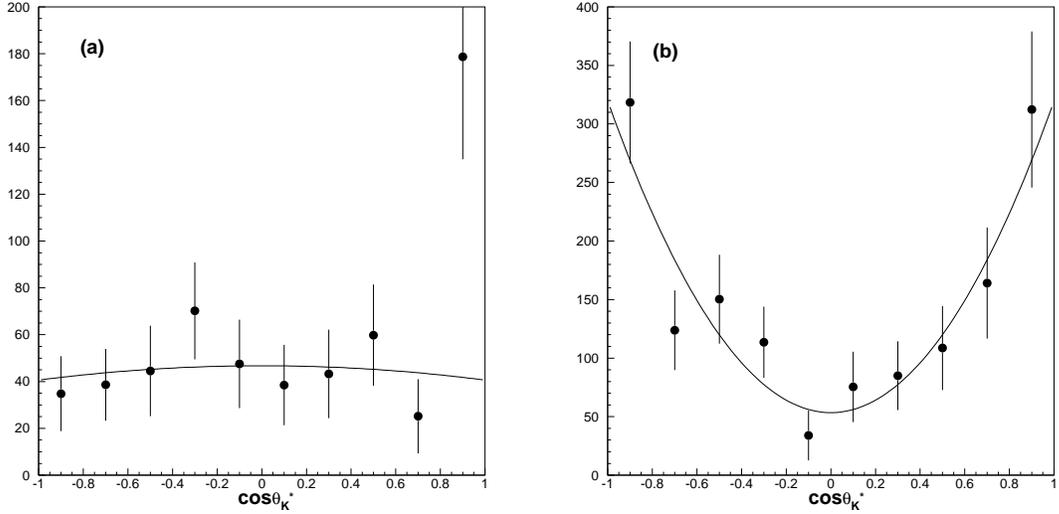}
\end{tabular}
\caption{The efficiency-corrected $\cos \theta _{K^*}$ distributions for $B^0\to {\chi}_{c1}K^{*0}(K^+\pi^-)$ decay 
mode fitted with $F(\cos \theta _{K^*}) 
= A+3B\cos ^2 \theta _{K^*}$ function for (a) background region (${\rm M_{K\pi}} < 0.8~{\rm GeV/c^2}$) and (b) 
signal region ($ 0.821~{\rm GeV/c^2} < {\rm M_{K\pi}} < 0.971~{\rm GeV/c^2}$).} \end{center}
\end{figure}

\begin{center}
{\small
\begin{table}
\begin{sideways}
Table 1
\end{sideways}
\begin{sideways}
The summary of the estimated branching fractions for $B\to {\chi}_{c1(2)}K(K^*)$ decay modes. The first (second) 
\end{sideways}
\begin{sideways}
{ quoted error is statistical (systematic). }
\end{sideways}
\begin{sideways}
\begin{tabular}{lccccc} \\ \hline
Mode                                & Efficiency  & Yield          & Br. Fraction ($\mathcal B$) & U.L.(90$\%$ C.L.)& Sign.\\ 
                                     & $\mathcal E (\%)$ &                & $\times 10^{-4}$ & $\times 10^{-4}$& ($\Sigma$)    \\ \hline
$B^+\to{\chi}_{c1}K^+ $                    & 25.4  & $645.7\pm26.9$   &$4.49\pm0.19\pm 0.53$  &     &37.4  \\        
$B^0\to{\chi}_{c1}K^0$                     & 18.5  & $127.0\pm11.8$     &$3.51\pm0.33\pm 0.45$  &     &15.3  \\ \hline 
$B^0\to{\chi}_{c1}K^{*0}(K^+\pi^-)$        & 17.4  & $209.9\pm22.7$    &$3.19\pm0.35\pm 0.73$ &     &12.1\\
$B^0\to{\chi}_{c1}K^{*0}(K^0\pi^0)$        & 4.2  &$ 5.0 \pm 4.8$       &$1.83\pm1.76\pm 0.70$ &     &1.1 \\ 
$B^0\to{\chi}_{c1}K^{*0}$                  &       &                  &$3.14\pm 0.34 \pm 0.72$  &     &\\ \hline  
$B^+\to{\chi}_{c1}K^{*+}(K^0\pi^+)$        & 11.4  & $53.6\pm 11.1$     &$3.61\pm0.75\pm 0.89$  &     &6.9\\ 
$B^+\to{\chi}_{c1}K^{*+}(K^+\pi^0)$        & 7.0 & $ 62.6 \pm 13.2$     &$4.73\pm0.99\pm 1.09$  &     &6.0\\ 
$B^+\to{\chi}_{c1}K^{*+}$                  &       &                  &$4.05 \pm 0.59\pm 0.95$  &     &\\ \hline      
$B^+\to{\chi}_{c2}K^+   $                  & 24.9  & $14.8\pm7.7$      &$0.16\pm0.08\pm 0.02$  &0.29     &2.1 \\  
$B^0\to{\chi}_{c2}K^0   $                  & 18.3  &$2.6\pm2.1$        &$0.11\pm0.09\pm0.02$ &0.26 &1.6\\ \hline     
$B^0\to{\chi}_{c2}K^{*0}(K^+\pi^-)$        & 16.5  & $11.7\pm9.4$      &$0.29 \pm 0.24\pm0.08$  &     &1.3 \\  
$B^0\to{\chi}_{c2}K^{*0}(K^0\pi^0)$        & 4.6  & $4.3 \pm 3.5$        &$2.24\pm1.82\pm 0.53$ &     &1.6 \\ 
$B^0\to{\chi}_{c2}K^{*0}$                  &       &                   &$0.32 \pm 0.23 \pm 0.08$ &0.71 & \\ \hline
$B^+\to{\chi}_{c2}K^{*+}(K^0\pi^+)$        & 11.2  & $5.2 \pm 5.4$       &$ 0.56\pm 0.58 \pm 0.13$&     &1.1\\ 
$B^+\to{\chi}_{c2}K^{*+}(K^+\pi^0)$        & 7.3   & $2.5 \pm 7.2$      &$0.28\pm0.82\pm0.24$  &     &0.3 \\ 
$B^+\to{\chi}_{c2}K^{*+}$                  &       &                  &$0.47 \pm 0.47\pm 0.13 $  &1.27 &\\ \hline  

\end{tabular}
\end{sideways}
\end{table}
} 
\end{center}
\begin{center}
\begin{table}
\begin{sideways}
Table 2
\end{sideways}
\begin{sideways}
{Summary of the multiplicative systematic errors in terms of percent ($\%$) for all modes. Here the columns list the individual sys-}
\end{sideways}
\begin{sideways}
{-tematic error components: Trk., Lep. and Had. are those associated with tracking, lepton and hadron efficiencies; $\gamma / \pi^0$ refers to }
\end{sideways}
\begin{sideways}
{$\gamma$ and $\pi^0$ reconstruction; $K_S^0$ refers to the $K_S^0$ selection efficiency; B.F. refers to the errors on secondary branching fractions; Inter.}
\end{sideways}
\begin{sideways}
{refers to the interference between $K^*$ signal and non-resonant $K\pi$ background; Bkg. Shape refers to different background shapes}
\end{sideways}
\begin{sideways}
{used in studying interference; Ang. refers to the  error due to the uncertainty of the angular distributions; Fit. Para. lists sensiti-}
\end{sideways}
\begin{sideways}
{-vity to the fit parameters; Bkg. Errors refers to  different fitting models for the $J/\psi$ inclusive background; and Eff.  Corr. lists}
\end{sideways}
\begin{sideways}
{errors associated with the efficiency correction.}
\end{sideways}
\begin{sideways}
\begin{tabular}{|l|lllll|l|c|c|c|c|c|c|c|} \hline 
       &\multicolumn{5}{c|}{Efficiency Uncertainty}&          B.F.   & Inter.    & Bkg.&Ang. &Fit  &Bkg.  & Eff.&Total\\ \cline{2-6} 
Modes        &Trk. &Lep.&Had.&$\gamma/~\pi^0$&$K^0_S$        & &&Shape & &Para.($\pm 1\sigma$) &Errors&Corr.& \\ 
                                   &$\%$&$\%$&$\%$ &$\%$  &$\%$  &$\%$ &$\%$ &$\%$&$\%$ & $\%$ & $\%$ & $\%$  &$\%$ \\ \hline 
$B^+\to{\chi}_{c1}K^+$             &3.7  &2.7 &$-$ &2  &$-$   &10.5 &$-$ &$-$ &$-$&$-$ &$0.8$     &$0.3$&11.7\\  
$B^0\to{\chi}_{c1}K^0$             &6.6 &2.7 &$-$ &2  &1     &10.5 & $-$ & $-$&$-$&$-$&$0.9$     &$0.3$&12.9  \\    
$B^0\to{\chi}_{c1}K^{*0}(K^+\pi^-)$ &5.2 &2.7 &1.9   &2  &$-$   &10.5 &15.0 &10.8& $5.1$&$-$ &$2.0$   &$0.3$ &22.9 \\  
$B^0\to{\chi}_{c1}K^{*0}(K^0\pi^0)$ &6.6&2.7 &$-$ &6&1     &10.5 &15.0 & 10.8 & $5.1$&$-$ &$30.0$  &$0.0$  &38.3 \\   
$B^+\to{\chi}_{c1}K^{*+}(K^0\pi^+)$ &8.1&2.7 &0.8   &2  &1     &10.5 &15.0 & 10.8 & $5.1$&$-$ &$7.0$    &$0.4$ &24.6 \\
$B^+\to{\chi}_{c1}K^{*+}(K^+\pi^0)$ &3.7 &2.7 &1.1   &6&$-$   &10.5 &15.0 &10.8 & $5.1$&$-$ &$0.8$    &$0.3$ &23.2  \\  
$B^+\to{\chi}_{c2}K^+$              &3.7 &2.7 &$-$ &2  &$-$   &8.5 &$-$ &$-$ &$-$&$-$&$8.1$      &$2.0$ &12.9  \\ 
$B^0\to{\chi}_{c2}K^0$              &6.6&2.7 &$-$ &2  &1     &8.5 & $-$ &$-$ &$-$&$-$&$7.7$   &$3.8$ &14.2 \\  
$B^0\to{\chi}_{c2}K^{*0}(K^+\pi^-)$ &5.2 &2.7 &1.9   &2  &$-$   &8.5 &15.0 & 10.8 & $5.1$&$-$ &$15.4$  &$1.7$ &26.9  \\  
$B^0\to{\chi}_{c2}K^{*0}(K^0\pi^0)$ &6.6&2.7 &$-$ &6&1     &8.5 &15.0 & 10.8 & $5.1$&$-$ &$4.7$   &$ 2.3$ & 23.6 \\ 
$B^+\to{\chi}_{c2}K^{*+}(K^0\pi^+)$ &8.1& 2.7 &0.8   &2  &1     &8.5 &15.0& 10.8 & $5.1$&3.8& $ 1.9$ &$ 1.9$ &23.2\\   
$B^+\to{\chi}_{c2}K^{*+}(K^+\pi^0)$ &3.7 &2.7 &1.1   &6&$-$   &8.5 &15.0 &10.8 & $5.1$&8.0& $84.0$ & $ 4.0$ &87.4\\ \hline 
\end{tabular}
\end{sideways}
\end{table}
\end{center}
\begin{table}
\begin{sideways}
Table 3
\end{sideways}
\begin{sideways}
{Comparison between this measurement and those at BaBar~\cite{b1,b2}. The first (second) quoted error is stat-}
\end{sideways}
\begin{sideways}
{-istical (systematic). }
\end{sideways}
\begin{sideways}
\begin{tabular}{|c|c|c|c|c|}\hline
Mode & \multicolumn{2}{c|}{Belle}& \multicolumn{2}{c|}{BaBar}\\\hline
&Br. Fraction $\times 10^{-4}$ & U.L.(90$\%$ C.L.) $\times 10^{-4}$ &Br. Fraction $\times 10^{-4}$& U.L.(90$\%$ C.L.) $\times 10^{-4}$ \\\hline
$B^+\to{\chi}_{c1}K^+ $ &$4.49\pm0.19\pm 0.53$  & &$5.79\pm0.26\pm0.65$  &\\
$B^0\to{\chi}_{c1}K^0$  &$3.51\pm0.33\pm 0.45$  & &$4.53\pm0.41\pm0.51$  & \\
$B^0\to{\chi}_{c1}K^{*0}$ & $3.14\pm 0.34 \pm 0.72$& & $3.27 \pm 0.42\pm 0.64$ & \\
$B^+\to{\chi}_{c1}K^{*+}$ &$4.05\pm 0.59\pm 0.95$ && $2.94\pm0.95\pm0.98$ & \\
$B^+\to{\chi}_{c2}K^+   $ &$0.16\pm0.08\pm 0.02$  &0.29 &$0.09\pm0.10\pm 0.11 $ & 0.30\\
$B^0\to{\chi}_{c2}K^0   $ &$0.11\pm0.09\pm0.02$ &0.26  &$0.21\pm0.11 \pm0.13 $ & 0.41 \\
$B^0\to{\chi}_{c2}K^{*0}$ &$0.32 \pm 0.23 \pm 0.08$ &0.71 & $0.14\pm 0.11 \pm 0.14 $& 0.36 \\
$B^+\to{\chi}_{c2}K^{*+}$ &$0.47 \pm 0.47\pm 0.13$ &1.27 & $-0.15 \pm 0.05 \pm 0.14 $ &0.12 \\\hline
\end{tabular}
\end{sideways}
\end{table}

\begin{thebibliography}{99}
\bibitem{belle01} 
K.~Abe {\it et al.} (Belle Collaboration),
Phys. Rev. Lett. {\bf 86}, 2509 (2001);
           {\bf 87},     091802 (2001);
Phys. Rev.  D {\bf 66},   071102(R) (2002);
{\bf 66},   032007 (2002);
B. Aubert {\it et al.}, (BaBar Collaboration)
Phys. Rev. Lett. {\bf 86}, 2515 (2001); 
{\bf 87}, 091801 (2001);
Phys. Rev.  D {\bf 66},   032003 (2002);
Phys. Rev. Lett. {\bf 89}, 201802 (2002).
 
\bibitem{suzuki}
M. Suzuki, Phys. Rev. D {\bf 66}, 037503 (2002).

\bibitem{wir85}
M. Bauer, B. Stech, M. Wirbel, Z. Phys. C {\bf 34}, 103 (1987).

\bibitem{chic0}
K.Abe, {\it et al.} (Belle Collaboration),
 Phys. Rev. Lett. {\bf 88}, 031802 (2002).

\bibitem{chic01}
B. Aubert, {\it et al.} (BaBar Collaboration),
 Phys. Rev. D {\bf 69}, 071103 (2004).

\bibitem{chic02}
K.Abe, {\it et al.} (Belle Collaboration),
 Phys. Rev. D {\bf 71}, 092003 (2005).

\bibitem{chic2}
K.Abe, {\it et al}. (Belle Collaboration),
 Phys. Rev. Lett. {\bf 89}, 011803 (2002). 

\bibitem{colan}
P. Colangelo, F.De Fazio and T.N. Pham, Phys. Lett. B {\bf { 542}}, 71 (2002).

\bibitem{belle}
A. Abashian {\it et al.} (Belle Collaboration), Nucl. Instr. and Meth. A {\bf 479}, 117 (2002).
\bibitem{kekb}
S. Kurokawa and E. Kikutani, Nucl. Instr. and Meth. A {\bf 499}, 1 (2003).

\bibitem{Fox78}
 G. Fox and S. Wolfram, Phys. Rev. Lett., {\bf 41} 1581 (1978).

\bibitem{klm}
A. Abashian {\it et al.}, Nucl. Instrum. Methods Phys. Res., Sect. A {\bf { 491}} 69 (2002).

\bibitem{ecl}
K. Hanagaki {\it et al.}, Nucl. Instrum. Methods Phys. Res., Sect. A {\bf {485}} 490 (2002).

\bibitem{evtgen}
 We use the EvtGen $B$-meson decay generator developed by the CLEO and
 the BaBar Collaborations, see: 
 http://www.slac.stanford.edu/$\sim$lange/EvtGen/.

\bibitem{geant}
 CERN program library long writeup W5013, CERN, (1993).

\bibitem{breit}
 J. Pisut and M. Roos, Nucl. Phys. B {\bf 6}, 325 (1968).

\bibitem{itoh}
 K. Abe {\it et al.}, (Belle Collaboration),
 Phys. Lett. B {\bf 538}, 11 (2002).

\bibitem{Hag}
S. Eidelman {\it et al.}, (Particle Data Group), Phys. Lett. B {\bf 592}, 1 (2004).

\bibitem{Jpsibabar}
B. Aubert {\it et al.}, (BaBar Collaboration),
Phys. Rev. Lett. {\bf 87}, 241801 (2001).


\bibitem{Feld} G.J. Feldman and R.D. Cousins,
 Phys. Rev. D {\bf 57}, 3873 (1998). 

\bibitem{b1} B. Aubert {\it et al.}, (BaBar Collaboration),
 Phys. Rev. Lett. {\bf 94}, 171801 (2005).

\bibitem{b2} B. Aubert {\it et al.}, (BaBar Collaboration),  Phys. Rev. Lett. {\bf 94}, 141801 (2005).
\end{thebibliography}
\end{document}